\documentclass[sigconf]{acmart}
\usepackage{url}
\usepackage{booktabs} 
\usepackage{algorithmic}
\floatname{algorithm}{Procedure}
\usepackage{colortbl}
\usepackage[ruled]{algorithm2e} 

\usepackage{subcaption}
\usepackage{multirow}
\usepackage{graphicx}
\usepackage{float}

\setcopyright{rightsretained}

\acmDOI{10.475/123_4}

\acmISBN{123-4567-2567/08/06}

\acmConference{Mobiquitous 2017}{Nov. 2017}{MELBOURNE, AU} 
\acmYear{2017}
\copyrightyear{2017}

\acmPrice{15.00}

\begin{document}

\title[Multi-Person Brain Activity Recognition]{Multi-Person Brain Activity Recognition via Comprehensive EEG Signal Analysis}

\author{Xiang Zhang}
\affiliation{%
  \institution{University of New South Wales}
  \city{Sydney, Australia} 
}\email{xiang.zhang3@student.unsw.edu.au}
\author{Lina Yao}
\affiliation{%
  \institution{University of New South Wales}
  \city{Sydney, Australia} 
}\email{lina.yao@unsw.edu.au}

\author{Dalin Zhang}
\affiliation{%
  \institution{University of New South Wales}
  \city{Sydney, Australia} 
}\email{dalin.zhang@student.unsw.edu.au}

\author{Xianzhi Wang}
\affiliation{%
  \institution{Singapore Management University}
  \city{Singapore} 
}\email{xzwang@smu.edu.sg}

\author{Quan Z. Sheng}
\affiliation{%
  \institution{Macquarie University}
  \city{Sydney, Australia} 
}\email{michael.sheng@mq.edu.au}

\author{Tao Gu}
\affiliation{%
  \institution{RMIT University}
  \city{Melbourne, Australia} 
}\email{tao.gu@rmit.edu.au}

\begin{abstract}
An electroencephalography (EEG) based brain activity recognition is a fundamental field of study for a number of significant applications such as intention prediction, appliance control, and neurological disease diagnosis in smart home and smart healthcare domains. Existing techniques mostly focus on binary brain activity recognition for a single person, which limits their deployment in wider and complex practical scenarios. Therefore, multi-person and multi-class brain activity recognition has obtained popularity recently. Another challenge faced by brain activity recognition is the low recognition accuracy due to the massive noises and the low signal-to-noise ratio in EEG signals. Moreover, the feature engineering in EEG processing is time-consuming and highly relies on the expert experience.
In this paper, we attempt to solve the above challenges by proposing an approach which has better EEG interpretation ability via raw Electroencephalography (EEG) signal analysis for multi-person and multi-class brain activity recognition. 
Specifically, we analyze inter-class and inter-person EEG signal characteristics, based on which to capture the discrepancy of inter-class EEG data. Then, we adopt an Autoencoder layer to automatically refine the raw EEG signals by eliminating various artifacts. We evaluate our approach on both a public and a local EEG datasets and conduct extensive experiments to explore the effect of several factors (such as normalization methods, training data size, and Autoencoder hidden neuron size) on the recognition results. The experimental results show that our approach achieves a high accuracy comparing to competitive state-of-the-art methods, indicating its potential in promoting future research on multi-person EEG recognition.

\end{abstract}


\maketitle

\section{INTRODUCTION} 
\label{sec:introduction}





Brain activity recognition is one of the most promising research areas over the last few years.
It  has 
the potential to revolutionize a wide range of applications such as ICU (Intensive Care Unit) monitoring \cite{osman2017diagnosing}, appliance control \cite{lee2013brain, zhang2017enhancing}, assisted living of disabled people and elderly people \cite{de2015combining}, and diagnosis of neurological diseases \cite{faust2015wavelet}. For instance, people with Amyotrophic Lateral Sclerosis (ALS) generally have only 
limited physical capacities and  
they are unable to communicate with the outer world, such as performing most daily activities, e.g., turning on/off a light. 
In such occasions, brain activity recognition can help interpret their demands and assist them to live more independently with dignity through the mind-control intelligence. Brain activities are mostly represented by Electroencephalography (EEG) signals, which record the voltage fluctuations of brain neurons with the electrodes placed on the scalp in a non-invasive way.

Although brain activity recognition has been widely investigated over the past decade, it still faces several challenges such as multi-person and multi-class classification.
First, despite several studies on multi-person EEG classification, e.g., \cite{djemal2016three} employed a LDA (linear discriminant analysis) classifier to classify two datasets with nine and three subjects, there still has space for improvement over the existing methods in terms of the classification accuracy (86.06\% and 93\% over the two datasets in \cite{djemal2016three}). 
Second, 
to the best of our knowledge, most existing applications that adopt EEG classification are for diseases diagnosis (such as epilepsy and Alzheimer's diseases), which requires only binary classification (normal or abnormal).
However, there exist various other deployment occasions (e.g., smart home and assisted living) that demand multi-class EEG classification. For instance, EEG-based assisting robots require more than two commands (such as walking straight, turning left/right, and raising/lowering hands) to complete assisted living tasks. Regarding this, only 
some preliminary research exists, such as \cite{wang2012multi}, which adopted SVM to classify a four-class EEG dataset and achieved the accuracy of 70\%.

In this paper, we propose a novel brain activity recognition approach to classifying the multi-person and multi-class EEG data. We analyze the similarity of EEG signals and calculates the correlation coefficients matrix in both inter-class and inter-person conditions. Then, on top of data similarity analysis, we extract EEG signal features by the Autoencoder algorithm, and finally feed the features into the XGBoost classifier to recognize categories of EEG data, with each category corresponding to one specific brain activity. The main contributions of this paper are summarized as follows:

\begin{itemize}
\item We present a novel brain activity recognition approach based on comprehensive EEG analysis. The proposed approach directly works on the raw EEG data, which enhances the ductility, relieves from EEG signal 
pre/post-processing, and decreases the need of human expertise.
\item We calculate the correlation coefficients matrix and measure the self-similarity and cross-similarity under both inter-class and inter-person conditions. Based on the similarity investigation, we propose three favorable conditions of multi-person and multi-class EEG classification.
\item We adopt the Autoencoder, an unsupervised neuron network algorithm, to refine EEG features. Moreover, we investigate the size of hidden-layer neurons to optimize the neurons size to optimize the classification accuracy.
\item We conduct an experiment to evaluate the proposed approach on a public EEG dataset (containing 560,000 samples from 20 subjects and 5 classes) and obtain the accuracy of 79.4\%. Our approach achieves around 10\% accuracy improvement compared with other popular EEG classification methods. 
\item We design a case study to evaluate the proposed approach on a local dataset which consists of 172,800 samples collected from 5 subjects and 6 classes. Our approach obtains the accuracy of 74.85\% and outperforms the result of the state-of-the-art methods.
\end{itemize}

The rest of this paper is organized as follows. Some existing studies related to this paper are 
introduced in Section~\ref{sec:related_work}. Section~\ref{sec:eeg_characteristic_analysis} investigates the EEG data characteristic and provides the EEG sample similarity intra-class and inter-class. Section~\ref{sec:methodology} describes the methodology details of the approach adopted in this paper. The experimental results and evaluation are presented in Section~\ref{sec:experiment}. A local experiment case study is introduced in Section~\ref{sec:case_study}.
Finally, we summarized the paper and highlight the future work in Section~\ref{sec:discussion_and_conclusion}.

\section{RELATED WORK} 
\label{sec:related_work}
Over the last decade, much attention has been drawn to brain data modeling, a crucial pathway to translating human brain activity into computer commands to realize Brain-Computer Interaction (BCI). BCI systems are an alternative way to allow paralyzed or severely muscular disordered patients to recover communication and control abilities, as well as to save scarce medical care resources. Recent research has also found its application in virtual reality \cite{wairagkar2016movement} and space applications \cite{rossini2009brain}.
As EEG signals are the most commonly used brain data for BCI system \cite{zhang2016sparse, zhang2014frequency, arvaneh2013optimizing}, significant efforts have been devoted to build accurate and effective models for EEG-based brain activity analysis \cite{yahya2015adaptive,santana2012introducing,bhattacharyya2014automatic,muller2017towards}.
 
\textbf{EEG Feature Representation Method.}$\indent$Feature representation of EEG raw data has great impact on classification accuracy due to the complexity and high dimensionality of EEG signals.
Vézard et al. \cite{vezard2015eeg} employed common spatial pattern (CSP) along with LDA to pre-process EEG data and obtained an accuracy of 71.59\% on binary alertness states. 
Meisheri et al. \cite{meisheri2016multiclass} exploited multi-class CSP (mCSP) combined with Self-Regulated Interval Type-2 Neuro-Fuzzy Inference System (SRIT2NFIS) classifier for four EEG-based motor imagery classes (movement imagination of left hand, right hand, both feet, and tongue) classification and achieved the accuracy of 54.63\%, which is significantly lower than the accuracy of binary classification.
Shiratori et al. \cite{shiratori2015three} achieved a similar accuracy of 56.7\% using mCSP coupled to the random forests for a three-class EEG-based motor imagery task. The autoregressive (AR) modeling approach, a widely used algorithm for EEG feature extraction, is also broadly combined with other feature extraction techniques to gain a better performance \cite{rahman2012comprehensive}. For example, \cite{zhang2016classification} investigated two methods EEG with AR and feature extraction combination: 1) AR model and approximate entropy, 2) AR model and wavelet packet decomposition. They employed SVM as the classifier and showed that AR can effectively improve classification performance. 
Duan et al. \cite{duan2016feature} introduced the Autoencoder method for feature extraction and obtained an accuracy of 86.69\%.

\textbf{EEG Multi-person Classification.}$\indent$ Multi-person EEG classification investigates mental signals from multiple participants, each of whom undergoing the same brain activities. It is the requirement of future ubiquitous application of EEG instruments to capture the underlying consistency and inter-subject variations among EEG patterns of different subjects. 
Kang et al. \cite{kang2014bayesian} presented a Bayesian CSP model with Indian Buffet process (IBP) to investigate the shared latent subspace across subjects for EEG classification. Their experiments on two EEG datasets containing five and nine subjects showed the superior performance of approximate 70\% accuracy. Djemal et al. \cite{djemal2016three} utilized two multi-person multi-class EEG datasets to validate sequential forward floating selection (SFFS) and a multi-class LDA algorithm. Eugster et al. \cite{eugster2014predicting} involved forty participants in their experiments to perform relevance judgment tasks. They also recorded the EEG signals for further classification research. Ji et al. \cite{ji2016eeg} investigated a 
dataset containing nine subjects for analyzing and evaluating a hybrid brain-computer interface.
 
\textbf{EEG Multi-class Classification.}$\indent$ Multi-class classification is a major challenge in EEG signal analysis, given
 that
 current EEG classification research is mostly focused on binary classification. Usually, an algorithm achieves only inferior performance when handling multi-classification than in handling binary classification. Anh et al. \cite{anh2016artificial} used Artificial Neural Network trained with output weight optimization back-propagation (OWO-BP) training scheme for dual- and triple-mental state classification problems. They got a classification accuracy of 95.36\% on dual mental state for triple classification problems, the algorithm performance fell off to 76.84\%. With the four-class problem, Olivier et al. \cite{olivier2016independent} got an accuracy of around 50\% when using a voting ensemble neural network classifier. 
Aiming at four-class EEG classification, Wang et al. \cite{wang2012multi} employed four preprocessing steps and a simple SVM classifier and got an average classification accuracy of 70\%.

In summary, differing from previous work, this paper proposes an \textit{Autoencoder+XGBoost} algorithm to address the multi-class multi-person EEG signal classification problem, which is a core challenge in applying brain activity recognition technologies to many important domains. The proposed algorithm engages the \textit{Autoencoder} for EEG feature representation to explore the relevant EEG features. Also, it emphasizes on the generalization over participants by solving an EEG classification problem with as much as five classes and taking twenty subjects. The present approach is supposed to improve the accuracy and practical feasibility of EEG classification.


\begin{table*}[ht]
\centering
\caption{Inter-class correlation coefficients matrix. The correlation coefficients matrix (upper left section) is the average of 20 correlation coefficients matrix separately from 20 subjects.}
\label{tab:c1}
\begin{tabular}{llllll|ccc}
\hline
\rowcolor[HTML]{C0C0C0} 
\textbf{Class}   & \textbf{0}      & \textbf{1}      & \textbf{2}      &\textbf{ 3}      &\textbf{ 4}      & \textbf{Self-similarity} & \textbf{Cross-similarity} & \textbf{Percentage difference} \\ \hline
0       & 0.4010 & 0.2855 & 0.4146 & 0.4787 & 0.3700 & 0.401           & 0.3872           & 3.44\%                \\
1       & 0.2855 & 0.5100 & 0.0689 & 0.0162 & 0.0546 & 0.51            & 0.1063           & 79.16\%               \\
2       & 0.4146 & 0.0689 & 0.4126 & 0.2632 & 0.3950 & 0.4126          & 0.2854           & 30.83\%               \\
3       & 0.4787 & 0.0162 & 0.2632 & 0.3062 & 0.2247 & 0.3062          & 0.2457           & 19.76\%               \\
4       & 0.3700 & 0.0546 & 0.3950 & 0.2247 & 0.3395 & 0.3395          & 0.3156           & 7.04\%                  \\ \hline
Range   & 0.1932 & 0.4938 & 0.3458 & 0.4625 & 0.3404 & 0.2038          & 0.2809           & 75.72\%               \\
Average & 0.3900 & 0.1870 & 0.3109 & 0.2578 & 0.2768 & 0.3939          & 0.2680           & 28.05\%               \\
STD     & 0.0631 & 0.1869 & 0.1334 & 0.1487 & 0.1255 & 0.0700          & 0.0932           & 27.33\%               \\ \hline
\end{tabular}
\end{table*}

\begin{table*}[ht]
\centering
\caption{Inter-person correlation coefficients matrix. STD denotes Standard Deviation, SS denotes Self-similarity, CS denotes Cross-similarity, and PD denotes Percentage Difference.}
\label{tab:c2}
\resizebox{\textwidth}{!}{\begin{tabular}{llll|lll|lll|lll|lll}
\hline
\rowcolor[HTML]{C0C0C0} 
 & \multicolumn{3}{c}{\textbf{Class 0}} & \multicolumn{3}{c}{\textbf{Class 1}} & \multicolumn{3}{c}{\textbf{Class 2}} & \multicolumn{3}{c}{\textbf{Class 3}} & \multicolumn{3}{c}{\textbf{Class 4}} \\ \hline
 \rowcolor[HTML]{C0C0C0} 
\textbf{subjects} & \textbf{SS} & \textbf{CS} & \textbf{PD} & \textbf{SS} &\textbf{ CS} & \textbf{PD} & \textbf{SS} & \textbf{CS} & \textbf{PD} & \textbf{SS} & \textbf{CS} & \textbf{PD} & \textbf{SS} & \textbf{CS} & \textbf{PD} \\ \hline
subject1 & 0.451 & 0.3934 & 12.77\% & 0.2936 & 0.1998 & 31.95\% & 0.3962 & 0.3449 & 12.95\% & 0.4023 & 0.1911 & 52.50\% & 0.5986 & 0.4375 & 26.91\% \\
subject2 & 0.3596 & 0.2064 & 42.60\% & 0.3591 & 0.1876 & 47.76\% & 0.5936 & 0.3927 & 33.84\% & 0.2354 & 0.2324 & 1.27\% & 0.3265 & 0.2225 & 31.85\% \\
subject3 & 0.51 & 0.3464 & 32.08\% & 0.3695 & 0.2949 & 20.19\% & 0.3979 & 0.3418 & 14.10\% & 0.4226 & 0.3702 & 12.40\% & 0.4931 & 0.4635 & 6.00\% \\
subject4 & 0.3196 & 0.1781 & 44.27\% & 0.4022 & 0.1604 & 60.12\% & 0.3362 & 0.2682 & 20.23\% & 0.4639 & 0.3905 & 15.82\% & 0.3695 & 0.2401 & 35.02\% \\
subject5 & 0.4127 & 0.2588 & 37.29\% & 0.3961 & 0.2904 & 26.69\% & 0.3128 & 0.2393 & 23.50\% & 0.4256 & 0.1889 & 55.62\% & 0.3958 & 0.3797 & 4.07\% \\
subject6 & 0.33 & 0.2924 & 11.39\% & 0.3869 & 0.3196 & 17.39\% & 0.3369 & 0.3281 & 2.61\% & 0.4523 & 0.1905 & 57.88\% & 0.4526 & 0.3321 & 26.62\% \\
subject7 & 0.4142 & 0.3613 & 12.77\% & 0.3559 & 0.342 & 3.91\% & 0.3959 & 0.3867 & 2.32\% & 0.4032 & 0.3874 & 3.92\% & 0.4862 & 0.2723 & 43.99\% \\
subject8 & 0.362 & 0.1784 & 50.72\% & 0.4281 & 0.2121 & 50.46\% & 0.4126 & 0.2368 & 42.61\% & 0.3523 & 0.1658 & 52.94\% & 0.4953 & 0.2438 & 50.78\% \\
subject9 & 0.324 & 0.2568 & 20.74\% & 0.3462 & 0.2987 & 13.72\% & 0.3399 & 0.3079 & 9.41\% & 0.3516 & 0.1984 & 43.57\% & 0.3986 & 0.177 & 55.59\% \\
subject10 & 0.335 & 0.1889 & 43.61\% & 0.3654 & 0.2089 & 42.83\% & 0.2654 & 0.2158 & 18.69\% & 0.3326 & 0.2102 & 36.80\% & 0.3395 & 0.2921 & 13.96\% \\
subject11 & 0.403 & 0.1969 & 51.14\% & 0.3326 & 0.2066 & 37.88\% & 0.3561 & 0.3173 & 10.90\% & 0.4133 & 0.1697 & 58.94\% & 0.5054 & 0.44 & 12.94\% \\
subject12 & 0.4596 & 0.2893 & 37.05\% & 0.4966 & 0.3702 & 25.45\% & 0.3326 & 0.2506 & 24.65\% & 0.4836 & 0.3545 & 26.70\% & 0.3968 & 0.3142 & 20.82\% \\
subject13 & 0.3956 & 0.2581 & 34.76\% & 0.4061 & 0.3795 & 6.55\% & 0.3965 & 0.3588 & 9.51\% & 0.3326 & 0.1776 & 46.60\% & 0.3598 & 0.3035 & 15.65\% \\
subject14 & 0.3001 & 0.299 & 0.37\% & 0.3164 & 0.2374 & 24.97\% & 0.4269 & 0.3763 & 11.85\% & 0.3856 & 0.1731 & 55.11\% & 0.4629 & 0.3281 & 29.12\% \\
subject15 & 0.3629 & 0.3423 & 5.68\% & 0.3901 & 0.2278 & 41.60\% & 0.7203 & 0.2428 & 66.29\% & 0.3623 & 0.3274 & 9.63\% & 0.3862 & 0.3303 & 14.47\% \\
subject16 & 0.3042 & 0.1403 & 53.88\% & 0.3901 & 0.3595 & 7.84\% & 0.4236 & 0.331 & 21.86\% & 0.4203 & 0.1634 & 61.12\% & 0.4206 & 0.3137 & 25.42\% \\
subject17 & 0.396 & 0.1761 & 55.53\% & 0.3001 & 0.2232 & 25.62\% & 0.6235 & 0.3579 & 42.60\% & 0.5109 & 0.198 & 61.24\% & 0.3339 & 0.2608 & 21.89\% \\
subject18 & 0.4253 & 0.3194 & 24.90\% & 0.3645 & 0.2286 & 37.28\% & 0.6825 & 0.222 & 67.47\% & 0.4236 & 0.3886 & 8.26\% & 0.4936 & 0.3017 & 38.88\% \\
subject19 & 0.5431 & 0.3059 & 43.68\% & 0.3526 & 0.2547 & 27.77\% & 0.4326 & 0.3394 & 21.54\% & 0.5632 & 0.3729 & 33.79\% & 0.4625 & 0.219 & 52.65\% \\
subject20 & 0.3964 & 0.3459 & 12.74\% & 0.3265 & 0.2849 & 12.74\% & 0.4025 & 0.3938 & 2.16\% & 0.3265 & 0.1873 & 42.63\% & 0.3976 & 0.2338 & 41.20\% \\ \hline
Min & 0.3001 & 0.1403 & 0.37\% & 0.2936 & 0.1604 & 3.91\% & 0.2654 & 0.2158 & 2.16\% & 0.2354 & 0.1634 & 1.27\% & 0.3265 & 0.177 & 4.07\% \\
Max & 0.5431 & 0.3934 & 55.53\% & 0.4966 & 0.3795 & 60.12\% & 0.7203 & 0.3938 & 67.47\% & 0.5632 & 0.3905 & 61.24\% & 0.5986 & 0.4635 & 55.59\% \\
Range & 0.2430 & 0.2531 & 55.16\% & 0.2030 & 0.2191 & 56.21\% & 0.4549 & 0.1780 & 65.31\% & 0.3278 & 0.2271 & 59.97\% & 0.2721 & 0.2865 & 51.53\% \\
Average & 0.3902 & 0.2667 & 31.40\% & 0.3689 & 0.2643 & 28.14\% & 0.4292 & 0.3126 & 22.96\% & 0.4032 & 0.2519 & 36.84\% & 0.4288 & 0.3053 & 28.39\% \\
STD & 0.0644 & 0.0723 & 0.1695 & 0.0456 & 0.0636 & 0.1518 & 0.1223 & 0.0589 & 0.1853 & 0.0717 & 0.0890 & 0.2066 & 0.0690 & 0.0759 & 0.1485\\ \hline
\end{tabular}
}
\end{table*}

\section{EEG CHARACTERISTIC ANALYSIS} 
\label{sec:eeg_characteristic_analysis}

To gain knowledge about EEG data characteristics and prepare for the further EEG classification, we quantify the similarity between EEG samples by calculating their Pearson correlation coefficients, using the following equation: 

$$\rho (A,B)=\frac{1}{\bar{N}-1}\sum_{\bar{i}=1}^{\bar{N}}(\frac{A_{\bar{i}}-\bar{\mu_{A}}}{\bar{\sigma _{A}}})(\frac{B_{\bar{i}}-\bar{\mu_{B}}}{\bar{\sigma _{B}}}), \bar{i}=1,2,\dots, \bar{N}$$
where $A$ and $B$ denote two EEG vector samples, each containing $\bar{N}$ elements. $\mu_{A}$ and $\sigma_{A}$ denote the mean and standard deviation of $A$. $\mu_{B}$ and $\sigma_{B}$ denote the mean and standard deviation of $B$.
The Person correlation coefficient 
is 
positively correlated with the similarity, and both are in the range of $[0,1]$.

We introduce two similarity concepts used in our measurement: self-similarity and cross-similarity. The self-similarity is defined by the similarity of EEG signals within the same EEG category while the cross-similarity is defined by the similarity of EEG signals of two different EEG categories. Both the self-similarity and cross-similarity are measured under two conditions: inter-class and inter-person, respectively.

\textbf{Inter-class measurement.}$\indent$Under the inter-class situation, we measure the correlation coefficient matrix for \textit{every specific subject} and calculate the average matrix by calculating the mean value of all the matrix. For example, there are 5 classes for the specific subject, we calculate a $5*5$ correlation coefficient matrix. In this matrix, $\rho_{\breve{i},\breve{j}}$ denotes the correlation coefficient between the samples of the class $\breve{i}$ and the samples of the class $\breve{j}$. The self-similarity indicates the similarity between two different samples from the same class. The cross-similarity indicates the average of similarity of each possible \textit{class pair} of samples belonging to the \textit{specific subject}. 

\textbf{Inter-person measurement.}$\indent$Under the inter-person situation, we measure the correlation coefficients matrix for \textit{every specific} class and then calculate the average matrix. The self-similarity
 indicates the similarity between two different samples from the same class of the same subject. The cross-similarity denotes the average of similarity of each possible \textit{subject pair} of samples belonging to the \textit{specific class}.

Table~\ref{tab:c1} shows the inter-class correlation coefficient matrix and the corresponding statistical self- and cross-similarity. The last column (PD) denotes the Percentage Difference between the self-similarity and cross-similarity. We can observe from the results that the self-similarity is always higher than the cross-similarity for all classes, meaning that the samples' intra-class cohesion is stronger than the inter-class cohesion. The percentage difference has a noticeable fluctuation, indicating  the varying intra-class cohesion over different class pairs. \textit{Class 1} is easier to be distinguished due to its highest percentage difference, while in contrast, \textit{class 0} and \textit{class 4} are difficult to be accurately classified. 

Similarly, Table~\ref{tab:c2} shows the inter-person correlation coefficient matrix and gives an alternative visualization of the results. Again, we find that, for each class, the self-similarity is higher than cross-similarity with varying percentage difference. The standard deviations of cross-similarity in the five classes are similar. This indicates 
the steady and even distribution of the dataset between different subjects and different classes.

The above analysis results basically satisfy our following hypothesis for multi-person multi-class classification: 1) the self-similarity is consistently higher than cross-similarity both under inter-class and inter-person conditions; 2) the higher inter-class percentage difference, the better classification results; 3) lower average percentage differences and standard deviations of the subjects result in the better classification performance under the inter-person condition. 


\section{METHODOLOGY} 
\label{sec:methodology}

\begin{figure*}[]
\includegraphics[width=\linewidth]{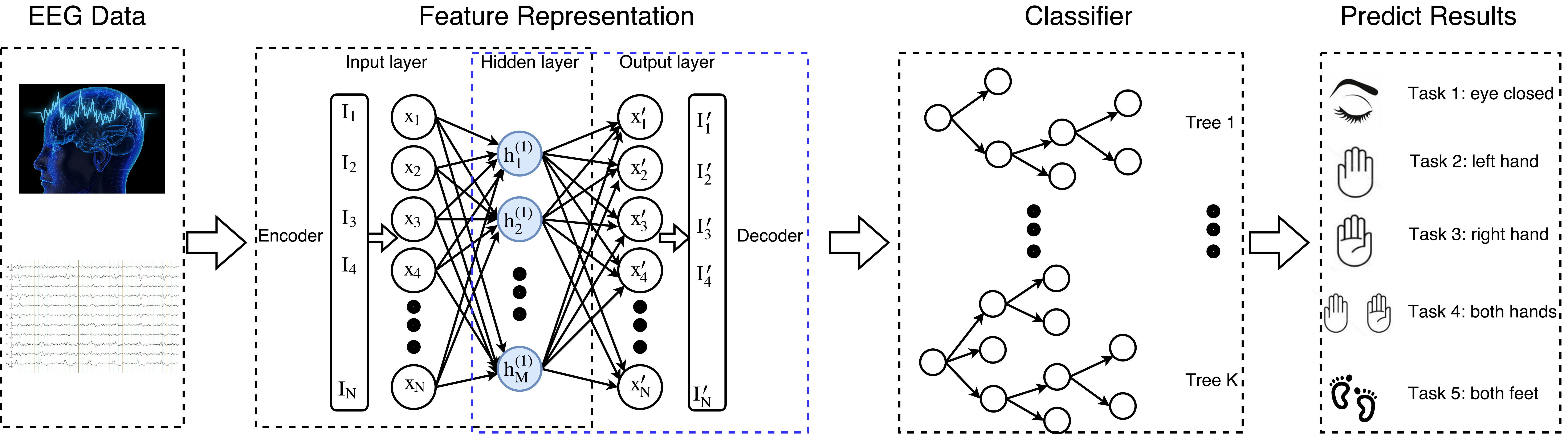}
\caption{The methodology flowchart. The collected EEG data flow into the Feature Representation component to seek for the appropriately representation and interpretation. $I_i$ and $I'_i$ separately indicate the input and output EEG data. $x_i$, $h_i$, and $x'_i$ indicate the neurons in the input layer, the hidden layer and the output layer, respectively. The learned feature representation $h$ will be sent to an XGBoost classifier with $K$ trees. The classifier's predict result is corresponding to the user's brain activity, which indicates the user's intention such as 
closing eye, moving left hand or moving right hand.}
\label{fig:overview}
\end{figure*}

In this section, we review the algorithm by first normalizing the input EEG data and then automatically explore the feature representation of the normalized data. At last, we adopt the XGBoost classifier to classify the trained features. The methodology flowchart is shown in Figure~\ref{fig:overview}.

\subsection{Normalization} 
\label{sub:normalization}
Normalization plays a crucial role in a knowledge discovery process for handling different units and scales of features. For instance, given one input feature ranges from 0 to 1 while another ranges from 0 to 100, the analysis results will be dominated by the latter feature. Generally, there are three widely used normalization methods: Min-Max Normalization, Unity Normalization, and Z-score Scaling (also called standardization). 

\textbf{Min-Max Normalization.}$\indent$ Min-Max Normalization projects all the elements in an vector to the range of $[0,1]$. This method maps features to the same range despite of their original means and standard deviations. The formula of Min-Max normalization is given below:
$$x_{new}=\frac{x-x_{min}}{x_{max}-x_{min}}$$
where $x_{min}$ and $x_{max}$ separately denotes the minimum and maximum in the feature $x$.

\textbf{Unity Normalization.}$\indent$ Unity Normalization re-scales the features by the percentage or the weight of each single element. It calculates the sum of all the elements and then divides each element by the sum. The equation is:
$$x_{new}=\frac{x }{ \sum x}$$
where $\sum x$ denotes the sum of feature $x$. Similar to Min-Max Normalization, the results of this method also belong to the range of $[0,1]$.

\textbf{Z-score Scaling.}$\indent$ Z-score Scaling forces features under normal Gaussian distribution (zero mean and unit variance), using the equation below:
$$x_{new}=\frac{x-\mu }{\sigma }$$
where $\mu$ denotes the expectation of feature $x$ and $\sigma$ denotes the standard deviation.

Depending on the feature characteristics of datasets, these 3 categories of normalization methods may lead to differed analysis results.

\subsection{Feature Representation} 
\label{sub:autoencoder}
To exploit the deeper correlationship between EEG signals, we adopt Autoencoder to have a better representation of EEG. The Autoencoder \cite{nguyen2015eeg} is an unsupervised machine learning algorithm that aims to explore a lower-dimensional representation of high-dimensional input data for \textit{dimensionality reduction}. In structure, Autoencoder is a multi-layer back propagation neural network that contains three types of layers: the input layer, the hidden layer, and the output layer. The procedure from the input layer to the hidden layer is called \textit{encoder} while the procedure from the hidden layer to the output layer is called \textit{decoder}. Both the encoder and the decoder yield a set of weights $W$ and biases $b$. 
Autoencoder is called either \textit{Basic Autoencoder} when there is only one hidden layer or \textit{Stacked Autoencoder} when there are more than one hidden layers. Based on our prior experiment experience, basic Autoencoder works better than stacked Autoencoder when dealing with EEG signals. Therefore, in this paper, we adopt the basic Autoencoder structure. 


Let $X=\{X_i|i= 1,2\cdots ,N\},X\in \mathbb{R}^N, X_i \in \mathbb{R}^d$ be the entire training data (unlabeled), where $X_i$ denotes the $i$-th sample, $N$ denotes the number of training samples, and $d$ denotes the number of elements in each sample. $h_{i}=\{h_{ij}|j=1,2,\cdots ,M\}, h_{i}\in \mathbb{R}^M$ represents the learned feature in the hidden layer for the $i$-th sample, where $M$ denotes the number of neural units in current layer (the number of elements in $h_{i}$). For simplicity, we use $x$ and $h$ to represent the input data and the data in the hidden layer, respectively. 
First, the encoder transforms the input data $x$ to the corresponding representation $h$ by the encoder weights $W_{en}$ and the encoder biases $b_{en}$:
$$h=W_{en}x+b_{en}$$
Then, the decoder transforms the hidden layer data $h$ to the output layer data $x'$ by the decoder weights $W_{de}$ and the decoder biases $b_{de}$:
$$x'=W_{de}h+b_{de}$$
The function of the decoder is to reconstruct the encoded feature $h$ and make the reconstructed data $x'$ as similar to the input data $x$ as possible. The discrepancy between $x$ and $x'$ is calculated by the MSE (mean squared error) cost function which is optimized by the RMSPropOptimizer.

In summary, training Autoencoder is the task of optimizing the parameters to achieve the minimum cost between the input $x$ and the reconstructed data $x'$. 
At last, the hidden layer data $h$ would contain the refined information. Such information can be regarded as representation of the input data, which is also the final outcome of Autoencoder. 
In above formulation, the dimension of the input data $x$ and the refined feature (the hidden layer data $h$) are $d$ and $M$, respectively. 
The function of the Autoencoder is either dimensionality reduction if $d>M$ or dimensionality ascent if $d<M$.


\subsection{Brain Activity Recognition} 
\label{sub:xgboost}
To recognize the brain activity based on the represented feature, in this section, we employ the XGBoost \cite{chen2016xgboost} classifier. 
XGBoost, also known as \textit{Extreme Gradient Boosting}, is a supervised scalable tree boosting algorithm
derived from the concept of \textit{Gradient Boosting Machine} \cite{friedman2001greedy}. Compared with gradient boosting algorithm, XGBoost proposes a more regularized model formalization to prevent over-fitting, with the engineering goal of pushing the limit of computation resources for boosted tree algorithms to achieve better performance.

Consider $n$ sample pairs $D=\{(x_{i'},y_{i'})\}$, $(\left|D\right|=n,x_{i'}\in \mathbb{R}^m,y_{i'}\in \mathbb{R})$ where $x_{i'}$ denotes a $m$-dimensional sample and $y_{i'}$ denotes the corresponding label. XGBoost aims to predict the label $\tilde{y_{i'}}$ of every given sample $x_{i'}$. 

The XGBoost model is the ensemble of a set of classification and regression trees (CART), each having its leaves and corresponding scores. The finial results of tree ensemble is the sum of all the individual trees. For a tree ensemble model of $K'$ trees, the predict output is:
$$\tilde{y_{i'}}=\sum_{k'=1}^{K}f_{k'}(x_{i'}), f_{k'}\in F$$
where $F$ is the space of all trees and $f_{k'}$ denotes a single tree.

The objective function of XGBoost includes loss function and regularization.
 The loss function evaluates the difference between each ground truth label $y_{i'}$ and the predict result $\tilde{y_{i'}}$. It can be chosen based on various conditions such as cross-entropy, logistic, and mean square error. 
The regularization part is the most outstanding contribution of XGBoost. It calculates the complexity of the model and a more complex structure brings larger penalty. 

The objective function is defined as:
\begin{equation}
\label{equ:object}
\Psi =\sum_{i}^nl(\tilde{y_{i'}},y_{i'})+\sum_{k'}^K\Omega (f_{k'})
\end{equation}
where $l(\tilde{y_{i'}},y_{i'})$ is the loss function and $\sum_{k'}\Omega (f_{k'})$ is the regularization term. The complexity of a single tree is calculated as
\begin{equation}
\label{equ:2}
\Omega (f_{k'})=\gamma T+\frac{1}{2}\lambda \|\omega\|^2
\end{equation}
where $T$ is the number of leaves in the tree, $\|\omega\|^2$ denotes the square of the L2-norm of the weights in the tree, $\gamma$ and $\lambda$ are the coefficients.
 The regularized objective helps deliver a model of simple structure and predictive functions. More specifically, the first term, $\Omega$, penalizes complex structures of the tree (fewer leaves lead to a smaller $\Omega$), while the second term  penalizes the overweights of individual trees in case of the overbalanced trees dominating the model. Moreover, the second term helps smooth the learned weights to avoid overfitting. 
\section{EXPERIMENT} 
\label{sec:experiment}
In this section, we evaluate the proposed approach on a public EEG dataset and report the results of 
our 
experimental studies. Firstly, we introduce the experimental setting and evaluation criterion. Then, we provide the classification results, 
followed by the analysis of influencing factors (e.g., normalization method, training data size, and neuron size in 
the 
Autoencoder hidden layer). 
Additional experiments are conducted to 
study 
the efficiency and robustness by 
comparing our approach with the state-of-the-art methods.

\subsection{Experimental Setting} 
\label{sec:experimental_setting}
We 
use the EEG data from PhysioNet eegmmidb (\textit{EEG
motor movement/imagery database}) database\footnote{\url{https://www.physionet.org/pn4/eegmmidb/}}, a widely used EEG database collected by the BCI2000 (Brain Computer Interface) instrumentation system\footnote{\url{http://www.schalklab.org/research/bci2000}} \cite{schalk2004bci2000}, to evaluate the proposed method. 
In particular, the data is collected by the BCI 2000 system, which owns 64 channels and an EEG data sampling rate of 160 Hz. 
During the collection of this database, the subject 
sits in front of one screen and performs the corresponding action as one target appears in different edges of the screen. According to the tasks, different annotations are labeled and can be downloaded from PhysioBank ATM \footnote{\url{https://www.physionet.org/cgi-bin/atm/ATM}}. The actions in different tasks are as follows:

\textit{Task 1}: The subject closes his or her eyes and keeps relax.

\textit{Task 2}: A target appears at 
the left side of the screen and then the subject focuses on the \textit{left hand} and \textit{imagines} he/she is opening and closing the left hand until the target disappears.

\textit{Task 3}: A target appears at the right side of the screen and then the subject focuses on the \textit{right hand} and \textit{imagines} he/she is opening and closing the right hand until the target disappears.

\textit{Task 4}: A target appears on the top of the screen, and the subject focuses on \textit{both hands} and \textit{imagines} he/she is opening and closing both hands until the target disappears.

\textit{Task 5}: A target appears on the bottom of the screen, and the subject focuses on \textit{both feet} and \textit{imagines} he/she is opening and closing both feet until the target disappears.

Specifically, we 
select 560,000 EEG samples from 20 subjects (28,000 samples each subject) for our experiments. Every sample is one vector which includes 64 elements corresponding to 64 channels. Each sample corresponding to one task (from task 1 to task 5 separately is \textit{eye closed}, \textit{focus on left hand}, \textit{focus on right hand}, \textit{focus on both hands} and \textit{focus on both feet}). Every task is labeled as one class, and there are totally 5 classes labels (from 0 to 4).

\subsection{Evaluation} 
\label{sub:evaluation}
Basic definitions related to classification problems include:
\begin{itemize}
\item True Positive (TP): the ground truth is positive and the prediction is positive; 
\item False Negative (FN): the ground truth is positive but the prediction is negative;
\item True Negative (TN): the ground truth is negative and the prediction is negative;
\item False Positive (FP): the ground truth is negative but the prediction is positive;
\end{itemize}
Based on these concepts, we define criteria to evaluate the performance of the classification results as follows:

\textbf{Accuracy.}$\indent $ The proportion of all correctly predicted samples. Accuracy is a measure of how good a model is. 
$$accuracy=\frac{TP+FN}{FP+FN+TP+TN}$$
The \textit{test error} used in this paper refers to the incorrectly predicted samples' proportion, which equals to 1 minus accuracy.

\textbf{Precision.}$\indent $ The proportion of all positive predictions that are correctly predicted. 
$$Precision=\frac{TP}{TP+FP}$$

\textbf{Recall.}$\indent $ The proportion of all real positive observations that are correctly predicted. 
$$Recall=\frac{TP}{TP+FN}$$

\textbf{F1 Score.}$\indent $ 
A `weighted average' of  precision and recall. The higher F1 score, the better the 
classification performance.
$$F1 \ Score =2\frac{precision*recall}{precision+recall}$$

\textbf{ROC.}$\indent $ The ROC (Receiver Operating Characteristic) curve describes the relationship between TPR (True Positive Rate) and FPR (False Positive Rate) at various threshold settings.


\textbf{AUC.}$\indent $ The AUC (Area Under the Curve) represents the area under the ROC curve. The value of AUC drops in the range $[0.5,1]$.
The higher the AUC, the better the classifier.



\subsection{Experiments and Results} 
\label{sec:experiments_and_results}

In our experiments, the Autoencoder model is trained by the training dataset then the testing dataset is fed into the trained Autoencoder model for feature extraction. 
The extracted features of the training dataset are used by the XGBoost classifier, which will be evaluated by the features of the testing dataset.
The number of neurons in the input and output layers in the Autoencoder model fixed at 64 (the input EEG data contains 64 dimensions), and the learning rate is set as $0.01$. Parameter tuning experience shows that 
Autoencoder performs better with more hidden layer neurons.
For XGBoost, we set the objective function as \textit{softmax} for multi-class classification through pre-experiment experience.
The parameters of XGBoost are selected based on the parameters tuning document\footnote{\url{https://github.com/dmlc/xgboost/blob/master/doc/parameter.md}}.
More specifically, we set the learning rate $\eta=0.7$,
the parameter related to the minimum loss reduction and the number of leaves $gamma=0$,
the maximum depth of a tree $max_{depth}=6$ (too large $max_{depth}$ may lead to overfitting), the subsampling ratio of training instance $subsample=0.9$ (to prevent overfitting), and $num_{class}=5$, since we have samples of 5 categories.
All the other parameters are set as default value. Without specific explanation, all the Autoencoder and XGBoost classifiers are taking above parameters setting. 

The hardware used in experiments is a GPU-accelerated machine with Nvidia Titan X pascal GPU, 768G memory, and 1.45TB PCIe based SSD. The training time is listed in related experiments, respectively.

\begin{table}[]
\centering
\caption{The confusion matrix of 5-class EEG classification and the performance evaluation (Precision, Recall, F1 score, and AUC)}
\label{tab:confusion}
\resizebox{0.5\textwidth}{!}{\begin{tabular}{lllllll|llll}
\hline
\rowcolor[HTML]{C0C0C0}
 &  & \multicolumn{5}{c|}{\textbf{Ground Truth}} & \multicolumn{3}{c}{\textbf{Evaluation}} &  \\ \hline
\multirow{6}{*}{\begin{tabular}[c]{@{}l@{}}Predict\\ Label\end{tabular}} &  & 0 & 1 & 2 & 3 & 4 & Precision & Recall & F1 Score & AUC \\
 & 0 & 3745 & 0 & 300 & 235 & 417 & 0.7973 & 0.7703 & 0.7836 & 0.9454 \\
 & 1 & 385 & 7857 & 515 & 445 & 488 & 0.8108 & 0.9219 & 0.8628 & 0.9572 \\
 & 2 & 245 & 174 & 3929 & 341 & 212 & 0.8017 & 0.7556 & 0.7780 & 0.9492 \\
 & 3 & 129 & 125 & 209 & 3304 & 153 & 0.8429 & 0.7294 & 0.7820 & 0.9506 \\
 & 4 & 358 & 367 & 247 & 205 & 3397 & 0.7427 & 0.7279 & 0.7352 & 0.9258 \\
total amount &  & 4862 & 8523 & 5200 & 4530 & 4667 & 3.9954 & 3.9051 & 3.9416 & 4.7282 \\
average &  &  &  &  &  &  & 0.7991 & 0.7810 & 0.7883 & 0.9456 \\ \hline
\end{tabular}
}
\end{table}

\begin{figure}
\includegraphics[width=\linewidth]{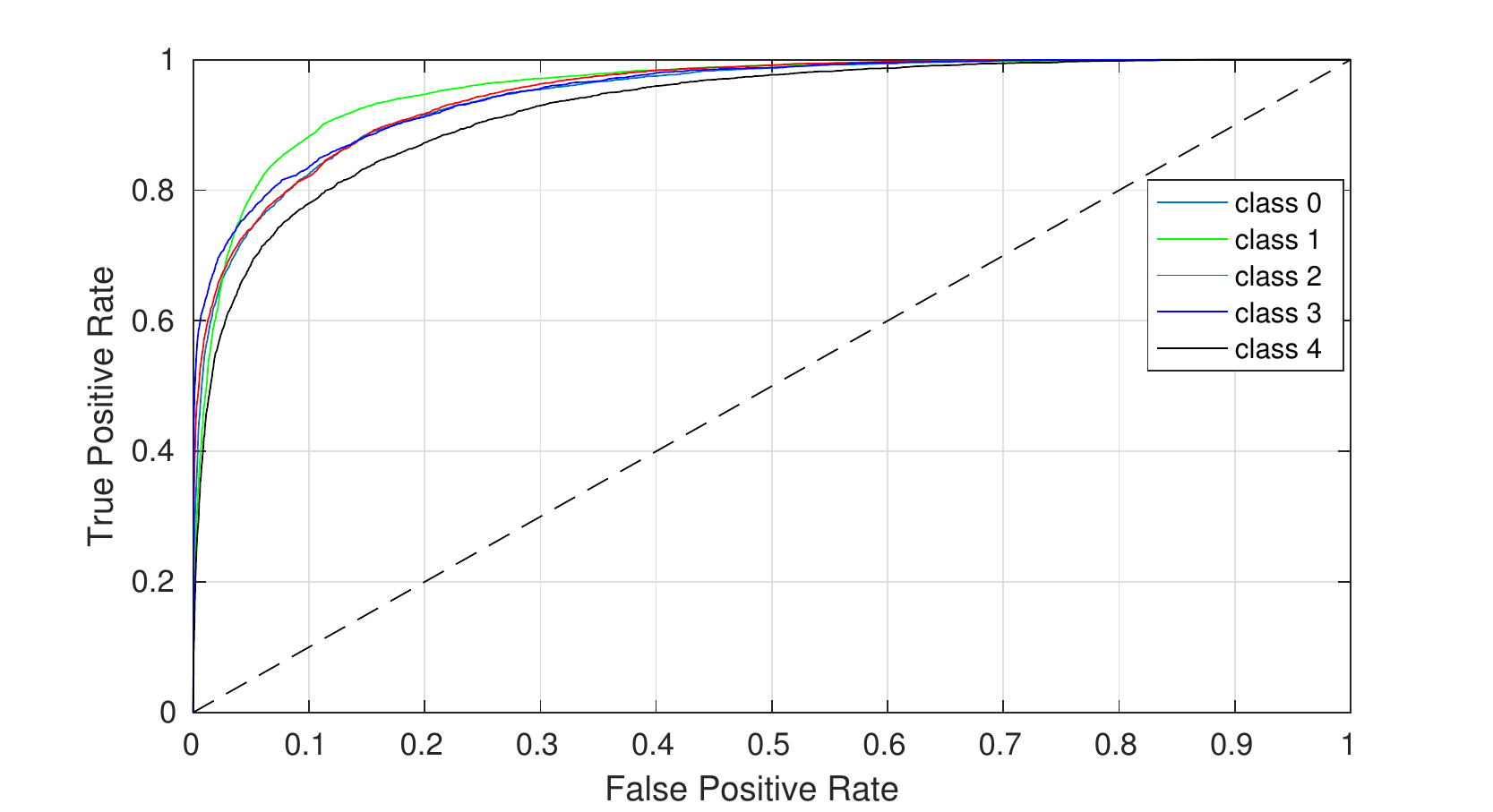}
\caption{ROC curve for 5-class classification by XGBoost. Five curves separately indicate the ROC curve of five classes. The dotted diagonal line denotes the random classifier where TPR=FPR. The closer the ROC curve to the upper left corner, the better performance the classifier has. 
It is clear to notice that the class 1 has the best classification performance.}
\label{fig:ROC}
\end{figure}

\textbf{Multi-person Multi-class EEG Classification.}$\indent$To evaluate the proposed approach, 560,000 EEG sample patches are utilized in this experiment. Each sample patch contains a feature vector of 64 dimensions and a ground truth label. The raw EEG data is normalized by the z-score scaling method and then randomly split into training dataset (532,000 samples) and testing dataset (28,000 samples). The representative features are extracted by 
Autoencoder with 121 hidden neurons and are input to the XGBoot classifier. The confusion matrix of the results is listed in Table~\ref{tab:confusion}. The classification accuracy of 28,000 testing samples (from 20 subjects and belong to 5 classes) is $0.794$. The average precision, recall, F1 score, and AUC are $0.7991$, $0.781$, $0.7883$, and $0.9456$, respectively. Among the evaluation standards, the \textit{class 1} obtains the highest precision, recall, F1 score, and AUC. This means that the \textit{class 1} samples have the most obvious divergence and are most distinguishable. On the contrary, \textit{class 4} is most confusing. This conclusion is highly consistent with our similarity analysis results in Section~\ref{sec:eeg_characteristic_analysis}. From the ROC curve, shown as Figure~\ref{fig:ROC}, we can deduce to the same conclusion. All the classes achieved the AUC higher than $0.92$, indicating that the classifier is steady and of high quality, according to the characteristics of AUC mentioned in Section~\ref{sub:evaluation}. 

\textbf{Effect of Normalization Method.}$\indent $The Autoencoder component regards the input data as the training target and calculates the discrepancy between them for error back propagation. This character of Autoencoder determines that the feature extraction quality and training cost are affected by the amplitude of the input data. In data pre-processing stage, the data values are directly related to the normalization method.
\begin{figure}
\includegraphics[width=\linewidth]{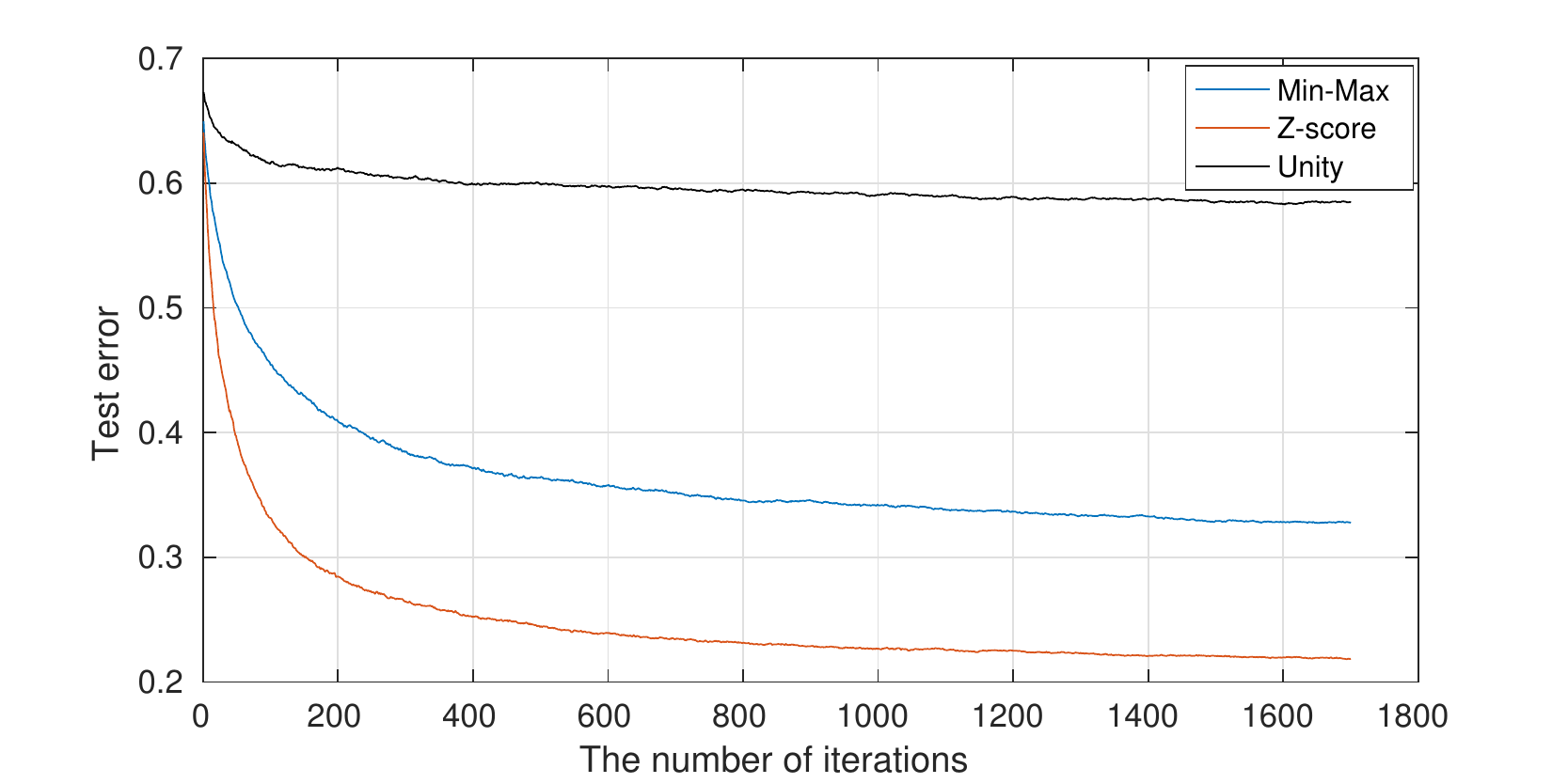}
\caption{The effect of normalization method. The three test error curves denote Min-Max, Z-score, and Unity normalization method, respectively.
 }
\label{fig:normalization_effect}
\end{figure}

To explore the impact of the normalization method, 560,000 EEG samples from 20 subjects are randomly split into a training dataset of 532,000 samples (95\% proportion) and a testing dataset of 28,000 samples (5\% proportion).
By setting 121 neurons for the hidden layer of 
Autoencoder, the XGBoost test error under three kinds of normalization methods is shown in Figure~\ref{fig:normalization_effect}. The figure shows that the z-score scaling normalization earns the lowest test error while the unity normalization obtains the highest test error. All the curves trend to convergence after 1,600 iterations. Without specific explanation, all the remaining experiments in this paper use the z-score scaling method.

\begin{figure}
\begin{minipage}[t]{\linewidth}
\includegraphics[width=\linewidth]{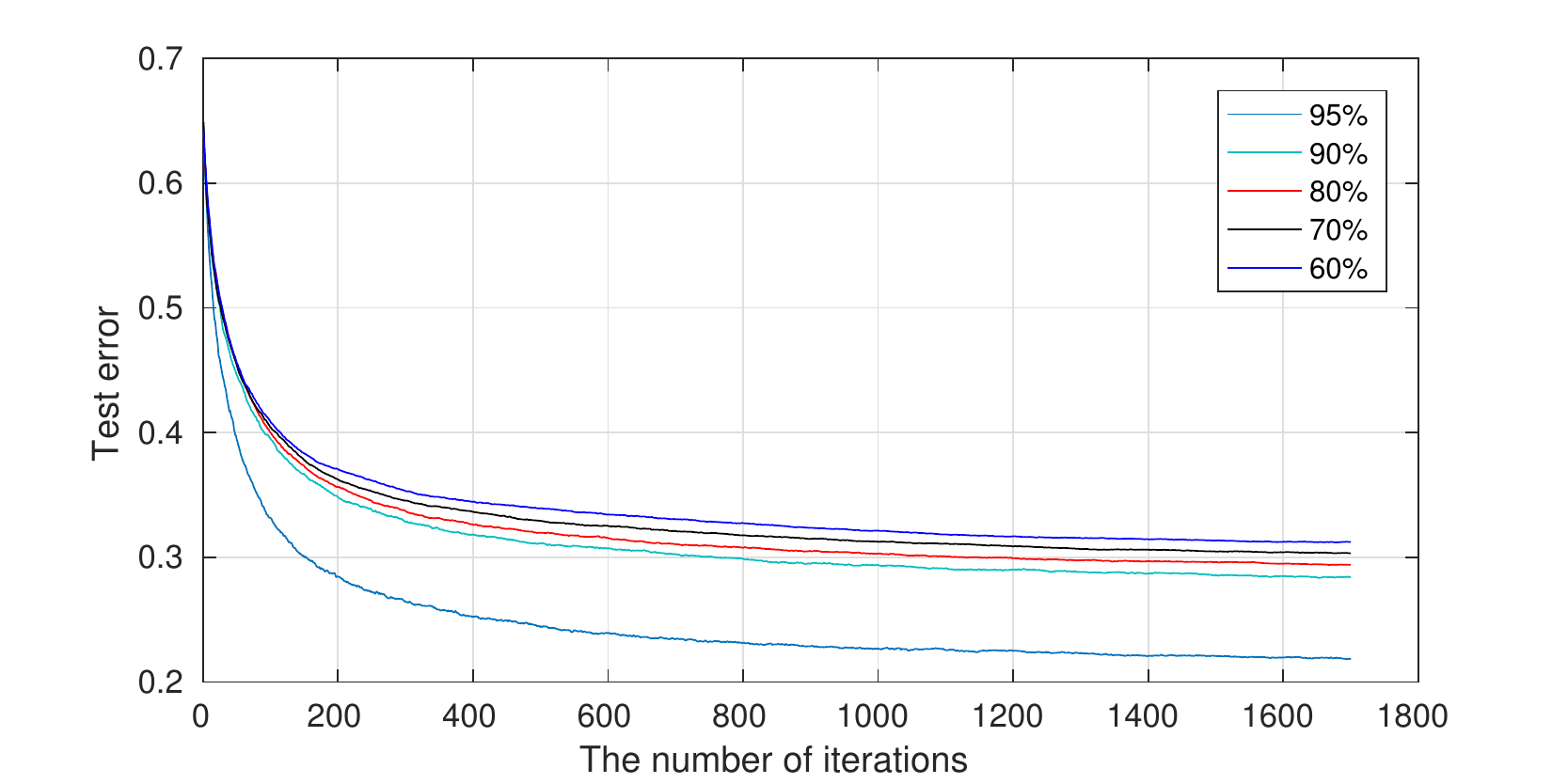}
\caption{The relationships between test error and the iterations under various training data proportions}
\label{fig:datasize_effect}
\end{minipage}
\begin{minipage}[t]{\linewidth}
\includegraphics[width=\linewidth]{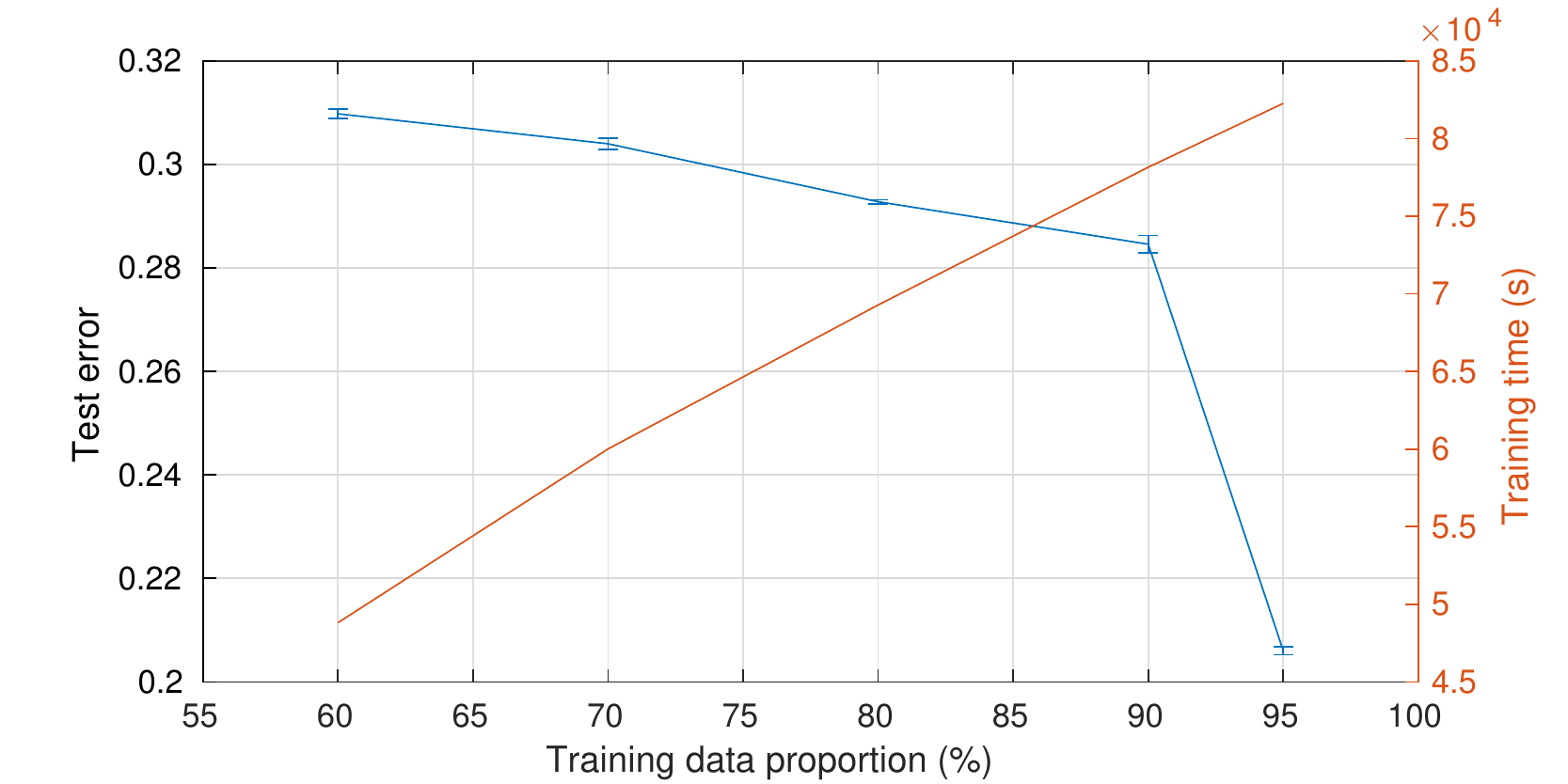}
\caption{The relationship between the test error with error bars, the training time and the training data proportion}
\label{fig:errorbar}
\end{minipage}
\vspace{-12.5mm}
\end{figure}

\textbf{Effect of Training Data Size.}$\indent$We explore in this section the relationship between the classification performance and the training data size. We design five experiments with the training data proportion of 60\%, 70\%, 80\%, 90\%, and 95\%, respectively.
Each experiment is repeated 5 times and the test error's error bar is shown in Figure~\ref{fig:errorbar}. The training time is positively correlated with the training data proportion. 
The test error arrives at the lowest point $0.206$, with an acceptable training time, while the proportion is 95\%. All the following experiments in this paper will take 95\% proportion. 

The relationships between test error and the iterations under various training data proportions are shown in Figure~\ref{fig:datasize_effect}. All the curves trend to convergence after 1,600 iterations and the higher proportion leads to lower test error.

\begin{table*}[]
\centering
\caption{Comparison of various classification methods. The first nine groups investigate the proper EEG data classifier and the last 7 groups illustrate the most efficient feature representation method.}
\label{tab:comparison}
\resizebox{\textwidth}{!}{\begin{tabular}{lllllllll}
\hline
\rowcolor[HTML]{C0C0C0}
\textbf{No.} & \textbf{1} & \textbf{2} & \textbf{3} & \textbf{4} & \textbf{5} & \textbf{6} & \textbf{7} & \textbf{8} \\ \hline
Classifier & SVM & RNN & LDA & RNN+SVM & CNN & DT & AdaBoost & RF \\
Acc & 0.3333 & 0.6104 & 0.3384 & 0.6134 & 0.5729 & 0.3345 & 0.3533 & 0.6805 \\ \hline
\rowcolor[HTML]{C0C0C0}
\textbf{No.} & \textbf{9} & \textbf{10} & \textbf{11} & \textbf{12} & \textbf{13} & \textbf{14} & \textbf{15} & \textbf{16} \\ \hline
Classifier & XGBoost & PCA+XGBoost & PCA+AE+XGBoost & EIG+AE+XGBoost & EIG+PCA+XGBoost & DWT+XGBoost & Stacked AE+XGBoost & AE+XGBoost \\
Acc & 0.7453 & 0.7902 & 0.6717 & 0.5125 & 0.6937 & 0.7221 & 0.7048 & \textbf{0.794} \\ \hline
\end{tabular}
}
\end{table*}

\begin{figure}
\includegraphics[width=\linewidth]{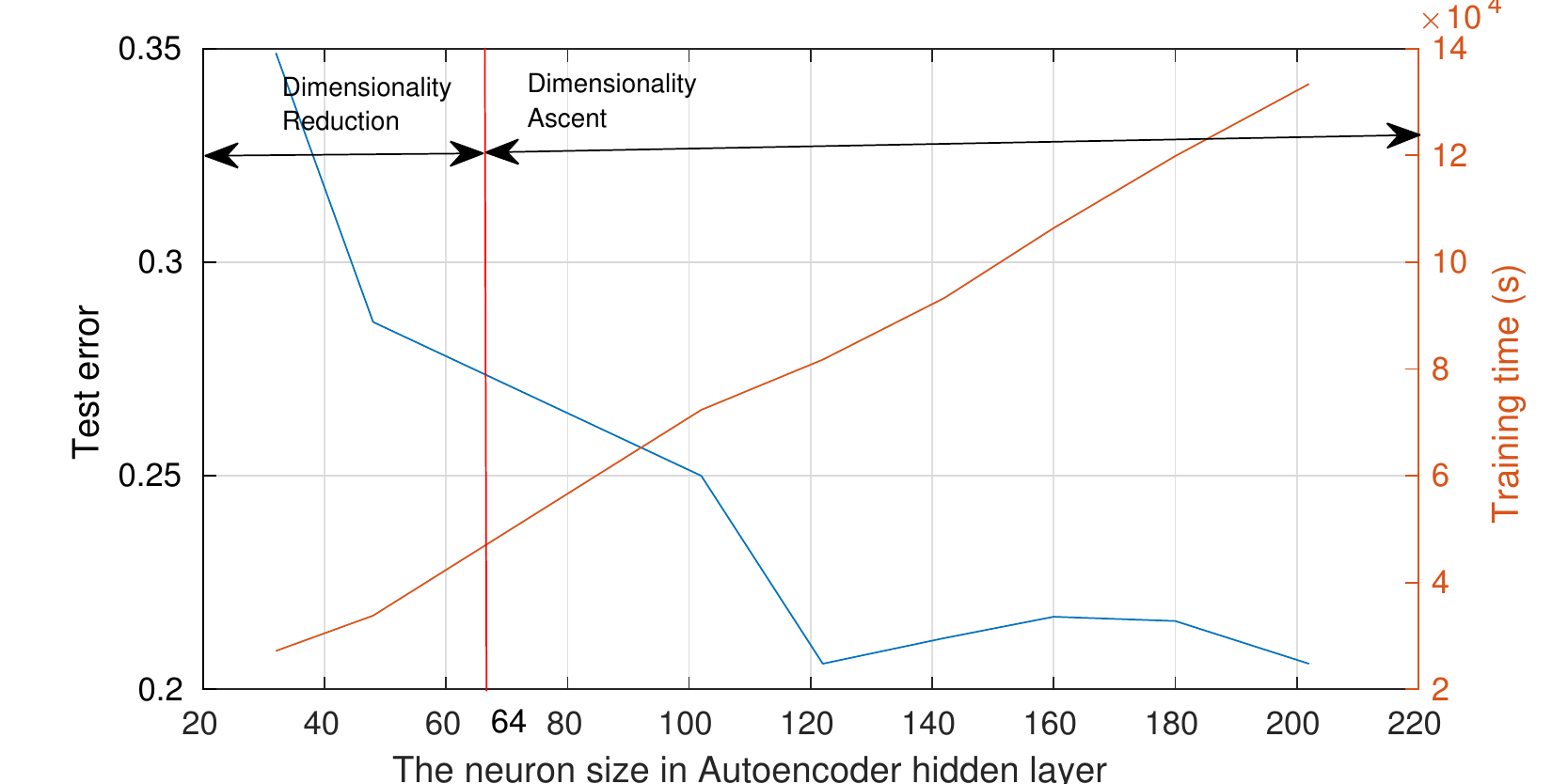}
\caption{The effect of neuron size in Autoencoder hidden layer. Since the input data is 64-dimension (marked as red line), the left part (smaller than 64) is dimensionality reduction area while the right part (larger than 64) is dimensionality ascent area.}
\label{fig:AE_hiddennumber}
\end{figure}

\textbf{Effect of Neuron Size in Autoencoder Hidden Layer.}$\indent$The neuron size in the hidden layer of 
Autoencoder indicates the number of dimensions of the extracted features. Thus it has great impact on the quality of feature extraction as well as the classification results. We design the experiment with the neuron size ranges from 30 to 200 and the experimental results (the test error and the training time) are shown in Figure~\ref{fig:AE_hiddennumber}. 

In the first stage (0-120), the test error keeps decreasing with the increase of 
the 
neuron size; in the second stage (larger than 120), the test error stands at around $0.21$ with slight fluctuation. The training time curve has a linear relationship with the neuron size on the whole. Although the gap between the test error curve and the training time curve arrives at the minimum around 100 neurons, the test error is still high. The test error reaches the bottom while the neuron size is 121, and the training time is acceptable at this point. Moreover, the test error curve keeps steady after 121. We set the hidden layer neuron size for all other experiments as 121. 

\begin{figure*}[h!]
    \centering
    \begin{subfigure}[b]{0.39\textwidth}
        \includegraphics[width=\textwidth]{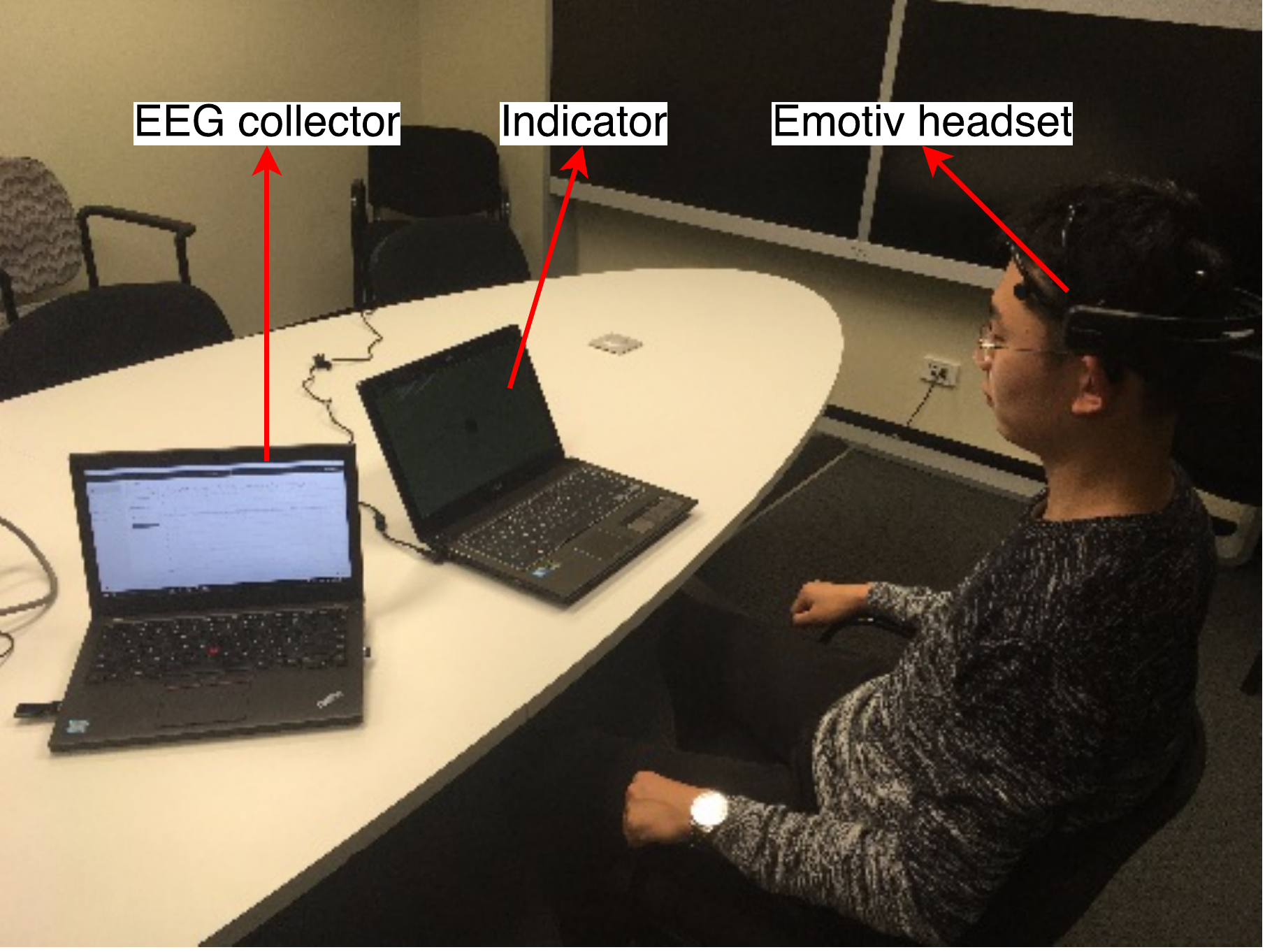}
        \caption{EEG collection}
        \label{fig:gull}
    \end{subfigure}%
    ~ 
    \begin{subfigure}[b]{0.5\textwidth}
        \includegraphics[width=\textwidth]{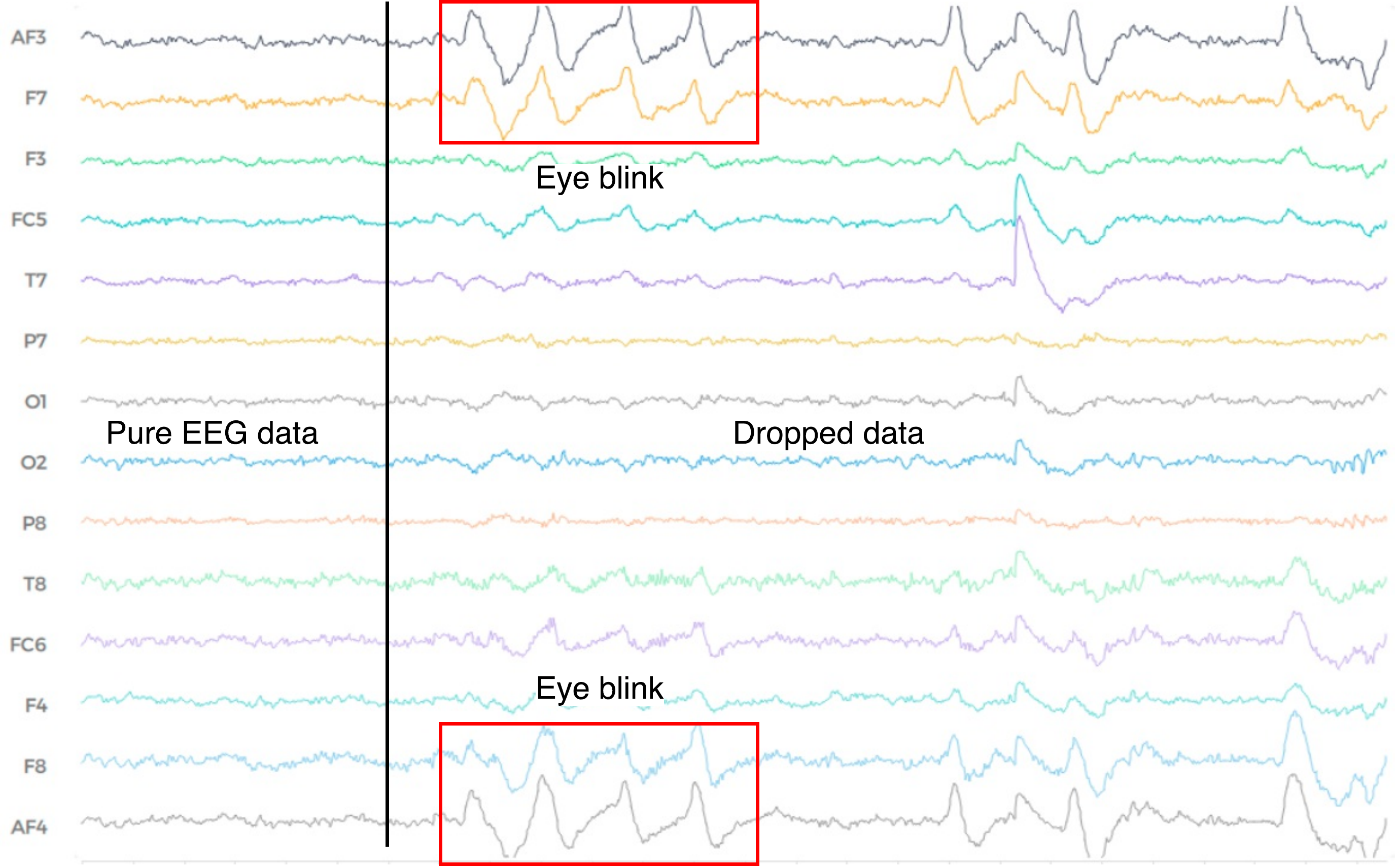}
        \caption{EEG raw data}
        \label{fig:tiger}
    \end{subfigure}
    \caption{EEG collection and the raw data
    . The pure EEG data is selected for recognition and the data, which is contaminated by eye blink and other noise, is not included in the local dataset (dropped).
    }
    \label{fig:EEG_collection}
\end{figure*}

\begin{table*}[]
\centering
\caption{Comparison of various classification methods over the case study dataset}
\label{tab:case_comparison}
\resizebox{\textwidth}{!}{
\begin{tabular}{lllllllll}
\hline
\rowcolor[HTML]{C0C0C0}
\textbf{No.} & \textbf{1} & \textbf{2} & \textbf{3} & \textbf{4} & \textbf{5} & \textbf{6} & \textbf{7} & \textbf{8} \\ \hline
Classifier & SVM & RNN & LDA & RNN+SVM & CNN & DT & AdaBoost & RF \\
Acc & 0.356 & 0.675 & 0.343 & 0.6312 & 0.5291 & 0.305 & 0.336 & 0.6609 \\ \hline
\rowcolor[HTML]{C0C0C0} \textbf{No.} & \textbf{9} & \textbf{10} & \textbf{11} & \textbf{12} & \textbf{13} & \textbf{14} & \textbf{15} & \textbf{16} \\ \hline
Classifier & XGBoost & PCA+XGBoost & PCA+AE+XGBoost & EIG+AE+XGBoost & EIG+PCA+XGBoost & DWT+XGBoost & StackedAE+XGBoost & AE+XGBoost \\ 
Acc & 0.6913 & 0.7225 & 0.6045 & 0.4951 & 0.6249 & 0.6703 & 0.6593 & {\bf0.7485} \\ \hline
\end{tabular}
}
\end{table*}
\subsection{Comparison} 
\label{sec:comparison}
In our approach, we employ XGBoost as the classifier to classify the refined EEG features yielded by Autoencoder. To demonstrate the efficiency of this method, in this section, we compare the proposed approach with several widely used classification methods. All the classifiers work on the same EEG dataset and their corresponding performance is listed in Table~\ref{tab:comparison}. 

In Table~\ref{tab:comparison}, LDA denotes Linear Discriminant Analysis; SVM denotes Support Vector Machine; RNN denotes (Recurrent Neuron Network) alongside LSTM denotes Long-Short Term Memory (kind of RNN architecture); AdaBoost denotes Adaptive Boosting; RF denotes Random Forest; DT denotes Decision Tree; EIG denotes the eigenvector-based dimensionality reduction method used in Eigenface recognition\footnote{\url{http://www.vision.jhu.edu/teaching/vision08/Handouts/case_study_pca1.pdf}}; PCA denotes Principal Components Analysis which is a commonly used dimensionality reduction method; DWT denotes Discrete Wavelet Transform, which is the wavelet transformation with the wavelets discretely sampled. The stacked Autoencoder contains 3 hidden layers with 100, 121, 100 neurons, respectively.
 
The comparison is divided into two aspects: the classifier and the feature representation method. At first, we classify our dataset separately by 9 commonly used sensing data classifier 
(e.g., SVM, RF, RNN, and CNN) to investigate the most suitable classifier for raw EEG data. Then 7 categories feature extraction method (e.g., PCA, AE, and DWT) are conducted to investigate the most appropriately EEG 
feature representation approach.
 The comparison results show that the XGBoost classifier outperforms its counterpart (without pre-processing and feature extraction) and obtains the accuracy of $0.74$, which means that XGBoost is more suitable to solve this problem. On the other hand, some feature extraction is positive to the classification whilst some are negative. Through the 
comparison, we 
find that 
Autoencoder (121 hidden neurons) achieves the highest multi-person classification accuracy as $\textbf{0.794}$.




\section{CASE STUDY} 
\label{sec:case_study}
In this section, to further demonstrate the feasibility of the proposed approach, we conduct a local experiment and present the classification result. At first, we design an EEG collection experiment. Secondly, we report the recognition classification results under the optimal hyper-parameters. Subsequently, we show the comparison between our approach and the state-of-the-art methods.

\subsection{Experimental Setting} 
\label{sec:section_name}
This experiment is carried on by 5 subjects (3 males and 2 females) aged from 24 to 30. During the experiment, the subject wearing the \textit{Emotiv Epoc+}\footnote{\url{https://www.emotiv.com/product/emotiv-epoc-14-channel-mobile-eeg/}} EEG collection headset, facing the computer screen and focus on the corresponding \textit{mark} which appears on the screen (shown in Figure~\ref{fig:EEG_collection}). The Emotiv Epoc+ contains 14 channels and the sampling rate is set as 128 Hz. The marks are shown on the screen and the corresponding brain activities and labels used in this paper are listed in Table~\ref{tab:label}. Summarily, this experiment contains 172,800 samples with 34,560 samples for each subject. 

\begin{table}[]
\centering
\caption{Mark in experiment and corresponding brain activity and label in case study}
\label{tab:label}
\resizebox{0.5\textwidth}{!}{
\begin{tabular}{lllllll}
\hline
\rowcolor[HTML]{C0C0C0}
\textbf{Mark} & \textbf{up arrow} & \textbf{down arrow} & \textbf{left arrow} & \textbf{right arrow} & \textbf{central cycle} & \textbf{close eye} \\ \hline
Brain Activity & upward & downward & leftward & rightward & center & relax \\
Label & 0 & 1 & 2 & 3 & 4 & 5\\ 
\hline
\end{tabular}
}
\end{table}
\begin{table}[]
\centering
\caption{The confusion Matrix and the 
evaluation}
\label{tab:case_confusion}
\resizebox{0.5\textwidth}{!}{
\begin{tabular}{llllllllllll}
\hline
\rowcolor[HTML]{C0C0C0}
\textbf{} & \multicolumn{7}{l}{\textbf{Ground truth}} & \multicolumn{4}{l}{\textbf{Evaluation}} \\ \hline
\multirow{7}{*}{Predicted label} &  & 0 & 1 & 2 & 3 & 4 & 5 & Precision & Recall & F1 & AUC \\
 & 0 & 2314 & 197 & 126 & 188 & 258 & 57 & 0.7369 & 0.7555 & 0.7461 & 0.8552 \\
 & 1 & 165 & 2271 & 153 & 171 & 214 & 75 & 0.7448 & 0.7407 & 0.7428 & 0.8395 \\
 & 2 & 131 & 176 & 2363 & 180 & 164 & 74 & 0.7652 & 0.7930 & 0.7788 & 0.8931 \\
 & 3 & 177 & 156 & 142 & 2179 & 216 & 85 & 0.7374 & 0.7230 & 0.7301 & 0.8695 \\
 & 4 & 191 & 190 & 118 & 219 & 2269 & 79 & 0.7401 & 0.6986 & 0.7187 & 0.8759 \\
 & 5 & 85 & 76 & 78 & 77 & 127 & 1539 & \textbf{0.7765} & 0.8062 & 0.7911 & 0.9125\\ \hline
\end{tabular}
}
\vspace{-5mm}
\end{table}


\subsection{Recognition Results and Comparison} 
\label{sec:recognition_results}
The dataset is divided into a training set (155,520 samples) and a testing set (17,280 samples). There are 9 mini-batches and the batch size is 17,280. All the other parameters are the same as listed in Section~\ref{sec:experiment}. The proposed approach achieves the 6-class classification accuracy of \textbf{0.7485}. The confusion matrix and evaluation is reported in Table~\ref{tab:case_confusion}. Clearly, the 5th class brain activity (eye closed and keep relax) has the highest precision and is the easiest activity to be recognized.

Subsequently, to demonstrate the efficiency of the proposed approach, we compare our method with the state-of-the-art methods and the results are shown in Table~\ref{tab:case_comparison}.

\section{CONCLUSION} 
\label{sec:discussion_and_conclusion}

In this paper, we have focused on  multi-class EEG signal classification based on EEG data that come from different subjects (multi-person). To achieve this goal, we aim at discovering the patterns in the discrepancy between different EEG classes with robustness over the difference between various subjects. 
Firstly, we analyze three widely used normalization methods in pre-processing stage. 
 Then, we feed the normalized EEG data into the Autoencoder and train the Autoencoder model. Autoencoder transforms the original 64-dimension features to 121-dimension features and essentially maps the data to a new feature space when meaningful features play a dominating role. Finally,  we evaluate our approach over an EEG dataset of 560,000 samples (\textit{belongs to 5 categories}) and achieve the accuracy of \textbf{0.794}. Compared with the accuracy of around $0.34$ achieved by traditional methods (e.g., SVM, AdaBoost, Decision Tree, and RNN), our results $0.794$ show significant improvement. Furthermore, we explore the effect of two factors (the training data size and the neuron size in Autoencoder hidden layer) on the training results. At last, we conduct a case study to gather 6 categories of brain activities and obtain the classification accuracy of 0.7485.

As part of our future work, we will build multi-view model of multi-class EEG signals to improve the classification performance. In particular, we plan to establish multiple models with each single model dealing with a single class.
Following this philosophy, the correlation between test sample and each model can be calculated in the test stage and the sample can be classified to the class with minimum correlation coefficient.
Besides, the work introduced in this paper only represents our preliminary study on exploring the common patterns of brain activities. 
Establishing a universal and efficient EEG classification model 
will be a major goal of our future research.

\bibliographystyle{ACM-Reference-Format}
\bibliography{sigproc} 

\end{document}